\documentclass{iopart}
\usepackage{graphicx,array,calc,tabularx,amsfonts,amssymb,rotating}
\usepackage{bm}

\begin{document}

\title[Lorentz violation in INTEGRAL $\gamma$-ray bursts]{Study of Lorentz violation in INTEGRAL gamma-ray bursts}
\author{Raphael Lamon$^1$, Nicolas Produit$^2$ and Frank Steiner$^1$}

\address{$^1$ Institut f\"ur Theoretische Physik, Universit\"at Ulm, Albert-Einstein-Allee 11, D-89069 Ulm, Germany}
\address{$^2$ INTEGRAL Science Data Center, Chemin d'Ecogia 16, CH-1290 Versoix, Switzerland}
\ead{raphael.lamon@uni-ulm.de}

\begin{abstract}
     We search for possible time lags caused by quantum gravitational (QG) effects using gamma-ray bursts (GRBs) detected by INTEGRAL. The advantage of this satellite is that we have at our disposal the energy and arrival time of every  detected single photon, which enhances the precision of the time resolution. We present a new method for seeking time lags in unbinned data using a maximum likelihood method and support our conclusions with Monte Carlo simulations. The analysis of the data yields a mass scale well below the Planck mass whose value may however increase if better statistics of GRBs were available. Furthermore, we disagree with previous studies in which a non-monotonic function of the redshift was used to perform a linear fit.
\end{abstract}

\maketitle

\section{\label{sec:Introduction}Introduction}
There is a general agreement that the classical space-time structure as described by the theory of general relativity will undergo drastic modifications at very small distances and very large energies due to quantum fluctuations. It is commonly argued that the relevant scales at which some new phenomena caused by quantum gravity (QG) occur are determined by a combination of Newton's constant $G$, Planck's constant $\hbar$ and the velocity of light in vacuo $c$, i.e. by the Planck length $l_P=\sqrt{\hbar G/c^3}\approx 1.6\cdot10^{-33}$ cm or equivalently the Planck energy $E_P$ or the Planck mass $M_P=E_P/c^2=\sqrt{\hbar c/G}\approx 1.2\cdot10^{19}$ $\mathrm{GeV}/c^2$. Until recently it was thought it would be almost impossible to detect the effects of such extremely short length scales or large energies.

Although a full quantum theory of gravity has not yet been established it has been realized that some generic predictions seem to emerge from the various approaches to a theory of QG. Assuming that QG possesses a well defined semiclassical limit which is obtained for weak gravitational fields and/or low energies, $E\ll E_P$, one can look for falsifiable predictions from semiclassical QG to first order in $E/E_P$.

One of the most striking predictions is a distortion of the photon dispersion relation
\begin{equation}\label{Esq}
E^2=p^2c^2+\alpha \frac{E^3}{E_P}+\mathcal{O}(E^4/E_P^2),
\end{equation}
where $E$, $p$ denote the photon energy and momentum, respectively, $\alpha$ is a model-dependent dimensionless parameter of order unity and $c$ is the (standard low-energy) velocity of light in vacuo. The non-standard dispersion relation \eref{Esq} leads to an energy-dependent velocity of light, $v=v(E)$, defined by the group velocity $v:=dE/dp$:
\begin{equation}\label{speedoflight}
v(E)=c\left(1+\alpha\frac{E}{E_P}\right)+\mathcal{O}\left((E/E_P)^2\right).
\end{equation}
The corrections to the velocity of light of the form \eref{speedoflight} could be interpreted as an explicit violation of Lorentz invariance at the Planck scale. For example, there may exist a preferred frame which is commonly chosen to be the frame that coincides with the rest frame of the cosmic microwave background radiation, implying that light would have a helicity-dependent velocity. In \cite{Kahniashvili:06} a more general dispersion relation with Lorentz symmetry breaking terms that depend explicitly on the helicity of the photon was studied and bounds on the QG scale from different astrophysical sources were given. However, observations of synchrotron radiation in the Crab nebula \cite{Jacobson:03} tend to rule out a helicity-dependent velocity of light.

Another possibility to understand a possible violation of Lorentz invariance, as proposed e.g. in non-critical string theory \cite{Ellis:92,Ellis:99}, string theory \cite{Kostelecky:89} or effective field theory approaches \cite{Myers:03}, is to interpret the energy-dependent velocity of light as $v(E)=c/n(E)$, where $n(E)$ is the refraction index of the non-trivial optical properties of the "foamy" structure of space-time caused by quantum fluctuations on short time and distance scales.

A very promising approach to QG is Loop Quantum Gravity (for reviews see \cite{Ashtekar:04,Bojowald:06,Rovelli:04,Smolin:04,Thiemann:01}), where modifications of the type \eref{Esq}-\eref{speedoflight} are present in the 2+1 dimensional theory \cite{Freidel:04} and where it is conjectured \cite{Smolin:02,Amelino:04} that the same will be true in QG in 3+1 dimensions. In this theory, the corrections are understood as indicating not a breaking of Lorentz invariance but rather a deformation of it. One assumes that the relativity of inertial frames is preserved, however, one requires that there be \textit{two} constant scales which are observer-independent: the standard velocity of light $c$ and the Planck length $l_P$ (or equivalently $E_P$). It has been shown that Lorentz invariant theories satisfying these requirements exist if the Lorentz transformations are treated not in the standard way, but are realized non-linearly when acting on energy and momentum eigenstates. Such theories are called Doubly or Deformed Special Relativity (DSR) \cite{Snyder:47,Amelino:01,Amelino:02,Amelino:03,Lukierski:91,Majid:91,Majid:93,Judes:03,Kowalski:02,Kowalski:05,Bruno:01,Magueijo:02,Magueijo:03,Smolin:06,Gambini:99,Girelli:04,Girelli:04:02}. Different realizations of DSR lead to a different energy dependence of the velocity of light. However, a common feature of all DSR models is that the velocity of light does not depend on helicity.

Note, however, that in \cite{Hossenfelder:07} it was argued that, in order to construct a quantum field theory that consistently incorporates DSR, it should not depend on extensive quantities like the total four momentum of particles but rather on intensive quantities like the fields' energy and momentum densities. This not only solves the soccer-ball problem but also changes drastically the predicted effects by many orders of magnitude. As shown there, the effect of the "new DSR" is about 57 orders of magnitude smaller than predicted by the "old DSR", thus making DSR hardly measurable.

In view of the fact that there exists a large variety of approaches to QG which lead to an energy-dependent velocity of light of the form \eref{speedoflight}, it seems worthwhile to seek experimental tests of \Eref{speedoflight}. It was pointed out that one powerful way to probe \Eref{speedoflight} may be provided by gamma-ray bursts (GRBs) \cite{Amelino:98,Scargle:06}. Several studies have been conducted using measurements of GRBs \cite{Ellis:03,Ellis:06,Bolmont:06}. GRBs are the most distant variable astrophysical sources of energetic photons detected by present experiments in the energy range from keV to GeV.

In this paper we shall consider the following form of the velocity of light
\begin{equation}\label{vofE}
v(E)=c\left(1\pm\frac{E}{Mc^2}\right),
\end{equation}
where we have put the constant $\alpha$ equal to $\pm 1$ and replaced the Planck mass $M_P$ by a QG mass $M$ to be determined or constrained by GRBs. Furthermore, we have neglected the higher order terms because the energy $E$ of the photons emitted by the available GRBs detected by INTEGRAL are much smaller than the expected energy scale $Mc^2$ representing the QG effects.

Light propagation from GRBs is not only determined by the velocity of light \eref{vofE} but is also affected by the cosmological expansion of the universe. Present observations are consistent with a nearly (spatially) flat universe. In the following we shall assume, for simplicity, an exactly flat universe described by the $\Lambda$CDM model consisting of baryonic matter (bar), cold dark matter (cdm) and a positive cosmological constant $\Lambda$, i.e. $\Omega_{\mathrm{tot}}=\Omega_{\mathrm{m}}+\Omega_{\Lambda}$ with $\Omega_{\mathrm{m}}:=\Omega_{\mathrm{bar}}+\Omega_{\mathrm{cdm}}=0.27$. The time delay between two photons with an energy difference  $\Delta E$ is then given by
\begin{equation}\label{deltatth}
\Delta t=\pm H_0^{-1}\frac{\Delta E}{Mc^2}\int_0^z\frac{dz}{\sqrt{\Omega_{\Lambda}+\Omega_{\mathrm{m}}(1+z)^3}},
\end{equation}
where $H_0=71\,\mathrm{km}\,\mathrm{s}^{-1}\,\mathrm{Mpc}^{-1}$ is the Hubble constant ($H_0^{-1}=13.77$ Gyr). Assuming an energy difference $\Delta E=300$ keV and a redshift $z=3$ we get a time lag of approximately $\Delta t=2\cdot10^{-5}$ s for $M=M_P$.

In spite of the fact that GRB signals are interesting for searching for QG effects, they are far from being perfect, mainly due to our lack of knowledge of the internal physical processes which are at the origin of the light emission. It is conceivable that photons of different energies are produced by different mechanisms within the GRB, thus narrowing the energy range in which a comparison between arrival times is possible.

In this paper we study possible time lags from GRB light curves detected by INTEGRAL. Contrary to previous studies \cite{Ellis:03,Ellis:06,Bolmont:06,Rodriguez:06:02,Boggs:04} where a binning in time and energy was used, we use unbinned data, i.e. we know the arrival time and energy of every single photon detected by the satellite. In order not to destroy this valuable piece of information we will not use wavelets as was done in the cases described above but rather use a new method.

The paper is organized as follows. In \sref{sec:INTEGRAL} we describe the relevant properties of INTEGRAL and in \sref{sec:FRED} our method to analyze the unbinned data. In \sref{sec:MC} we present the results of the Monte Carlo simulations. \Sref{sec:GRB}  contains the results using GRBs detected by INTEGRAL, and we finish with the conclusions in \sref{sec:conclusion}.

\section{INTEGRAL satellite}\label{sec:INTEGRAL}
INTEGRAL \cite{Winkler:03} is a mission of the European  
Space Agency (ESA) devoted to gamma ray astronomy.
It features a coded mask instrument ISGRI \cite{Lebrun:03}. This instrument enables us to measure for each photon
in the energy range 15 keV to 1 MeV the arrival time with a  
precision of $6\cdot10^{-5}$ s as well as the energy with a precision of 10\%.

The detector has a dead time
of about 25\%. This dead time is a function of the incoming  
rate and can vary during a GRB. The dead time is measured internally  
by the instrument and is given as a mean dead time over 8 seconds independently for 6  
parts of the detector. It can be corrected statistically
in weighing each incoming photon by $1/(1-\mathrm{dead}\; \mathrm{time})$ with the  
corresponding time slice and detector part dead time.
If the rate exceeds telemetry capabilities a data gap is created in  
wich the dead time is 100\%. In this case it cannot be  
statistically corrected
and we have a hole in the data versus time. This unfortunately  
happens frequently during very intense GRBs.

The instrument also registers an important rate from the  
background due to diffuse photons from the sky, internal
radioactivity of the instrument and flux from sources present in the  
field of view. This background rate varies with time but not  
perceptibly  during the typical time scale of a GRB. We have two  ways of predicting this background. Before and after the GRB the background can be measured as the full rate registered by the  
instrument. During the GRB, the pixels that are in the shadow of the mask for the  
direction of the GRB register only the background photons
of the GRB. The illuminated pixels register this background as well  
as the flux from the GRBs. Statistically the rate from the GRB can be  
computed by properly weighed
subtraction. As, most of the time, the GRBs are in the partially coded  
field of view, the number of pixels available for background  
measurement is bigger than the number of pixels
seeing the source.

The fraction of a pixel that is illuminated by the GRB (so called PIF  
value) can be calculated with the knowledge of the coordinate of the  
GRB and the knowledge of the attitude of the
instrument. We are not able to determine individually if a photon comes  
from the GRB or the background, but the PIF can be used
to properly weigh its probability to come from the GRB. For  
example, a light curve can be built by using only pixels that are  
fully illuminated by the source and removing
the constant rate measured by the completely opaque pixels.

\section{Description of the analysis method}\label{sec:FRED}
\begin{figure}[ht!]
\begin{center}
\includegraphics[width=7.5cm]{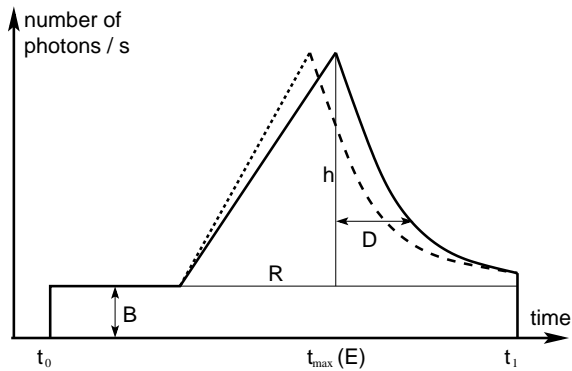}
 \caption{Sketch of a typical light curve of a GRB for a given energy interval. The curve is parameterized by five parameters: $ B $ is the background level, $ R $ the duration of the rise, $ h $ the height above the background, $ D $ the decay time for $\exp(-t/D)$ and  $\kappa$ describes the magnitude of the dependence on the energy of the distribution $f$, $ t_{\mathrm{max}}=P+\kappa\cdot E$, where $P$ is the time when the intensity reaches a maximum and $E$ is the photon energy. The area under the curve must be one, so that one parameter, e.g. $B$ , is fixed by this condition. The dashed line shows a distribution for another energy interval that is shifted by an amount of $\Delta t=\kappa\cdot\Delta E$ sketching the shift in time due to quantum gravitational effects. This shift is usually much smaller than the other parameters.}\label{fig:FRED}
\end{center}
\end{figure}

The majority of the GRBs seems to follow a pattern called Fast Raise and Exponential Decay (FRED). In order to model a GRB light curve, we parameterize it with five parameters and call the resulting probability distribution $f=f(t_i,E_i;P,B,R,D,\kappa,h)$ (see \fref{fig:FRED}). We suppose that a set of measured parameters $t_i$ and $E_i$ came from the probability density function $f$. We use the method of maximum likelihood, which consists of finding the set of values $\hat{P}$, $\hat{B}$, $\hat{R}$, $\hat{D}$, $\hat{\kappa}$ and $\hat{h}$, which maximizes the joint probability distribution for all data, given by
\begin{equation}\label{maxlikelihood}
\mathcal{F}(P,B,R,D,\kappa,h) =\prod_i f(t_i,E_i;P,B,R,D,\kappa,h)
\end{equation}
together with the constraint
\begin{equation}\label{constraint}
\int_{t_0}^{t_1} dt' \, f(t',E_i;P,B,R,D,\kappa,h) = 1,
\end{equation}
where $\mathcal{F}$ is the likelihood function and the integral runs between $t_0$ and $t_1$ as shown in \fref{fig:FRED}. In fact, the condition (\ref{constraint}) that the integral over time be equal to one reduces the degrees of freedom for $f$ and $\mathcal{F}$ by one. For example, $B$ can be chosen to be fixed by this condition, so we can think of $f$ and $\mathcal{F}$ as not depending on $B$. However, for clarity we write the $B$-term dependence for both functions.

It is easier to search for the parameters that maximize $\ln \mathcal{F}$, as the products on the right hand side of \Eref{maxlikelihood} is now a sum. To find these parameters, we use a multidimensional unconstrained nonlinear minimization where we minimize the function $-\ln \mathcal{F}$.

\Fref{fig:FRED} shows a typical light curve of a GRB. We always choose time intervals so that such a sketch can be found. However, in order to avoid wrong results, we also take account for other possibilities when for example $R>t_{\mathrm{max}}(E)-t_0$ or $t_1<t_{\mathrm{max}}(E)$.

\section{\label{sec:MC}Monte Carlo simulations}

\begin{figure}[ht!]
\begin{center}
\includegraphics[width=6.4cm]{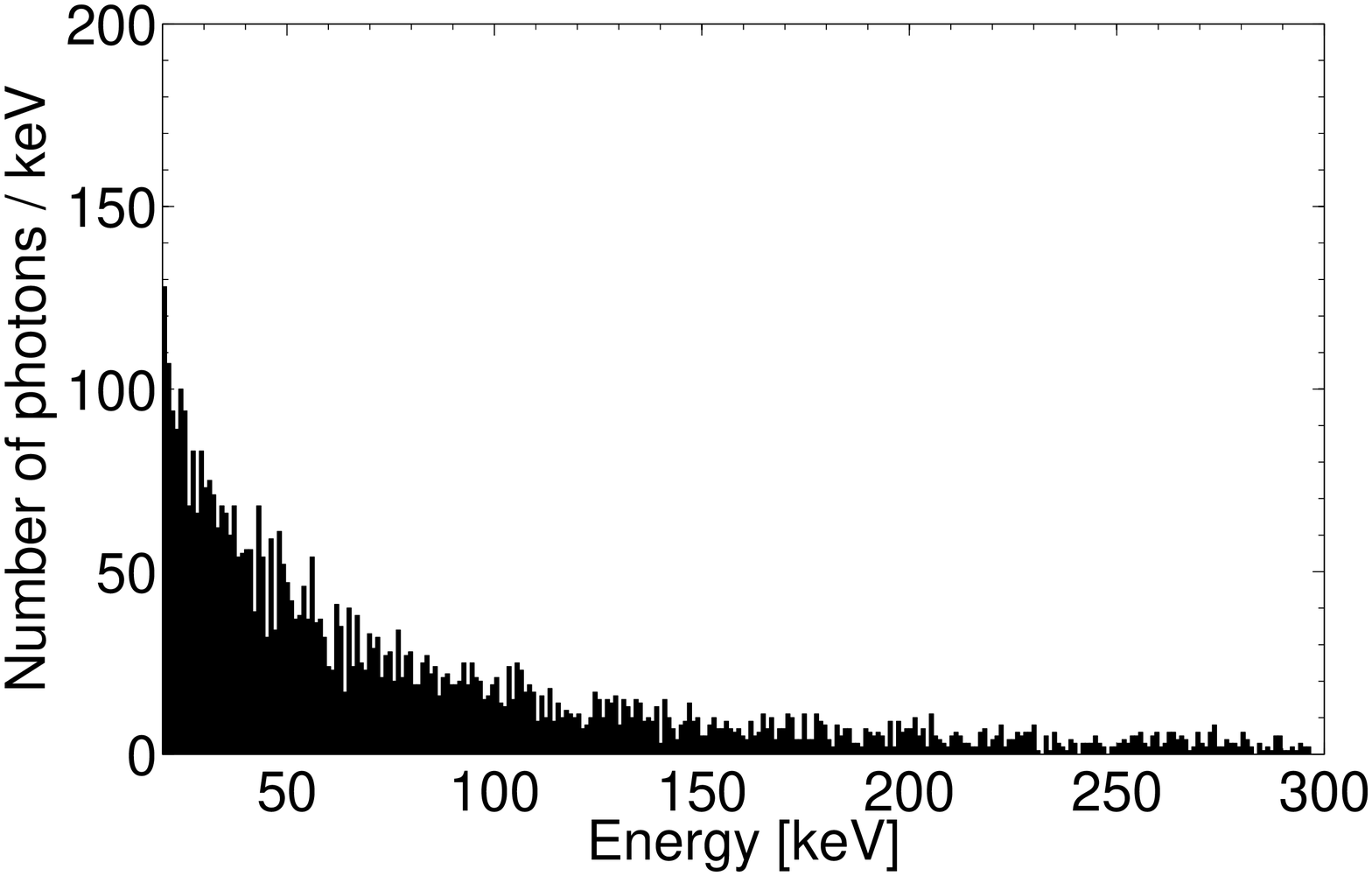}\includegraphics[width=6.4cm]{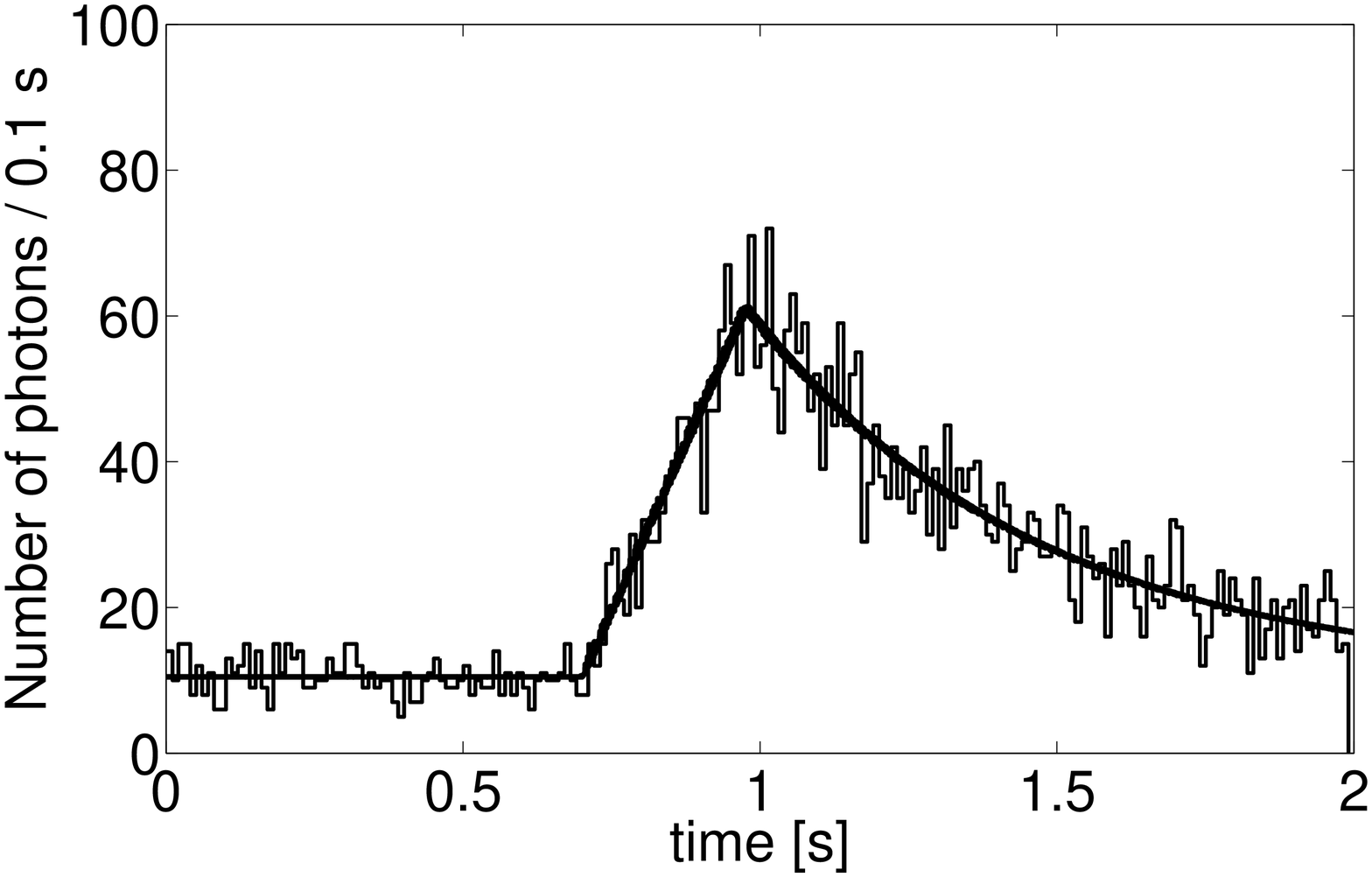}
\caption{Left panel: example of a  photon distribution as a function of the energy. The energy ranges from 20 keV to 300 keV according to the Band function (\ref{BAND}), the total photon number is 5000. Right panel: Example of a simulated GRB for $\kappa=-10^{-5}$ s/keV, $P=1$ s, $R=0.3$ s, $h=50$ $\mathrm{s}^{-1}$ and $D=0.5$ s with a total photon number of 5000. In order to be able to compare the fit with the GRB, we require that the areas under both curves be equal, so that the parameter $B$ is recovered. The overlaid curve is the FRED function with fitted parameters for a photon of energy 0. As the parameter $\kappa$ is very close to 0 this curve represents well the family of FRED curves of the problem.}
\label{fig:Energy}
\end{center}
\end{figure}

The maximum shift in time due to quantum gravity is expected to be of the order of $2\cdot10^{-5}$ s, which is smaller by a factor of three than the time resolution of INTEGRAL. Therefore, it is at first highly questionable whether such  time differences can be measured, not to speak of the results gotten from unbinned data. In order to get a better feeling of the behavior of the likelihood, we performed Monte Carlo simulations with a total number of photons ranging from 500 to unrealistic 300'000. First, we created $N$ events $i$ with energy $E_i$ distributed according to a typical GRB event. That is, a typical energy distribution for the photons of GRBs follows the pattern of the so-called Band function \cite{Band:93} given by the following equation:
\begin{eqnarray}\label{BAND}
N_E(E)&=&A\left( \frac{E}{100\;\mathrm{keV}}\right) ^{\alpha}\exp\left( -\frac{E}{E_0}\right) ,\nonumber\\
&& \quad\quad\quad\quad\quad\quad\quad\quad\quad\quad\quad(\alpha-\beta)E_0\geq E,\nonumber\\
&=&A\left[ \frac{(\alpha-\beta)E_0}{100\;\mathrm{keV}}\right] ^{\alpha-\beta}\left( \frac{E}{100\;\mathrm{keV}}\right) ^{\beta}\exp(\beta-\alpha),\nonumber\\
&& \quad\quad\quad\quad\quad\quad\quad\quad\quad\quad\quad(\alpha-\beta)E_0\leq E,
\end{eqnarray}
where we choose typical values for the parameters, i.e. $\alpha=-1$, $\beta=-2.5$ and $E_0=200$ keV (see left panel of \fref{fig:Energy}).

With this energy distribution, we created arrival times for each photon according to the FRED distribution $f$. In addition, because the time resolution of INTEGRAL is $6.1\cdot 10^{-5}$ s, we perturbed the arrival time of each photon with a Gaussian distribution with a deviation of $6.1\cdot 10^{-5}$ s. The Monte Carlo simulations were done with $\kappa=-10^{-5}$ s/keV, $P=1$ s, $R=0.3$ s, $h=50$ $\mathrm{s}^{-1}$ and $D=0.5$ s. Remember that $\Delta t=\kappa\cdot \Delta E$, so that a value for $\kappa$ of $10^{-5}$ s/keV represents a maximum time delay of $\sim 3\cdot10^{-3}$ s,  which is well longer than the expected time delay due to quantum gravitational effects.

\Fref{fig:Energy} (right panel) gives an example of a simulated GRB for parameters as described above. The histogram shows a typical simulation of a GRB using a FRED distribution, while the black line shows the solution of the minimization of \Eref{maxlikelihood}. This curve is defined by $\hat{P}=0.983$, $\hat{\kappa}=-4.95\cdot10^{-5}$ s/keV, $\hat{R}=0.276$ s,   $\hat{h}=48$ $\mathrm{s}^{-1}$ and   $\hat{D}= 0.485$ s. Except the value $\kappa$ which is five times too big, the other values are easily recovered by the minimization of \Eref{maxlikelihood}. However, the Monte Carlo simulations have a tendency to underestimate the parameters. As can be seen in \tref{table:resultsMC}, except the mean value of $D$ for $N=500$, all values are too low for small $N$. Note that, apart from $\kappa$, $R$ is not well estimated and has therefore a big deviation.

\setlength\extrarowheight{3pt}
 \begin{table}[h!]
    \caption{Results of 200 Monte Carlo simulations for each value of $N$}\label{table:resultsMC}\lineup
\begin{indented}
\item[]
\begin{tabular}{lllllll}
  \hline
$N$ & & $P$ [s] & $\kappa$ [s/keV] & $ R$ [s] & $h$ [$\mathrm{s}^{-1}$] & $D$ [s]\\
\hline
& $\bar{\mu}$ & 0.95 &	$-4.7\cdot10^{-5}$  & 0.17 &	4.57 &	0.65\\
\raisebox{1.5ex}[-1.5ex]{500} & $\sigma$ & 0.10 &\m	$4.0\cdot10^{-4}$ & 0.21 &	1.4 &	0.31\\
\hline
&$\bar{\mu}$ & 0.97 &	$-6.5\cdot10^{-7}$ & 0.19 & 	4.8 &	0.59\\
\raisebox{1.5ex}[-1.5ex]{1000} & $\sigma$ & $0.08$&\m$	2.7\cdot10^{-4}$& 0.21	& 1.11 &	0.24\\
\hline
& $\bar{\mu}$ & 0.99 &	$-2.3\cdot10^{-5}$ & 	0.19 &	5.1 &	0.51\\
\raisebox{1.5ex}[-1.5ex]{2000} & $\sigma$ & $0.03$ & \m$1.6\cdot10^{-4}$&	0.23 &	0.52 &	0.09\\
\hline
& $\bar{\mu}$ & 0.99 &	$-3.3\cdot10^{-6}$ &	0.17 &	5.0 &	0.50 \\
\raisebox{1.5ex}[-1.5ex]{5000} & $\sigma$ &0.02&	\m$8.6\cdot10^{-5}$ &	0.25 &	0.28 &	0.03\\
\hline
& $\bar{\mu}$ & 1.0 &	$-2.0\cdot10^{-5}$&0.17 &	5.0 &	0.50\\
\raisebox{1.5ex}[-1.5ex]{10000} & $\sigma$ &0.01 &	\m$6.8\cdot10^{-5}$&0.25 &	0.2 &	0.02\\
\hline
& $\bar{\mu}$ & 1.00 &	$-1.0\cdot10^{-5}$	& 0.30 &	5.00 &	0.50\\
\raisebox{1.5ex}[-1.5ex]{$3\cdot10^5$} & $\sigma$ & 0.002	& \m$ 1.2\cdot10^{-5} $&$	0.002$&$	0.04$&0.004\\
\hline

\end{tabular}
\end{indented}
\end{table}

From \tref{table:resultsMC} it should be clear that even with $3\cdot10^5$ photons it is not possible to get a trustful result for that small a value $\kappa$. Recall that $\kappa=10^{-5}$ s/keV is about a factor 100 larger than the expected time lags caused by quantum gravitational effects. A crude way of evaluating the statistics necessary for a convincing measurement is to make the assumption that the FRED distribution may be approximated by a Gaussian distribution. This distribution is obtained by minimizing the error of $N$ independent measurements, where the single parameters are $\bar{\mu}$ and $\sigma$. The error of a single measurement is given by $\sigma/\sqrt{N}$, so that if we want to reach a precision of $\Delta t=10^{-5}$ s with a burst lasting one second, we need $10^{10}$ photons.

\begin{figure}[ht!]
\begin{center}
\includegraphics[width=6.4cm]{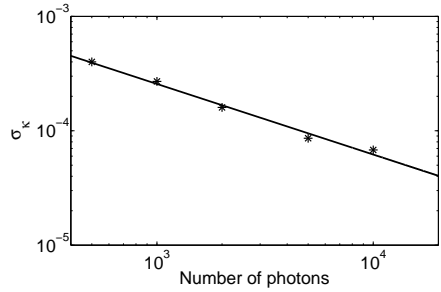}
\caption{Standard deviation $\sigma_\kappa$ for $\kappa$ as a function of the number of photons $N$ . The data points are shown in \tref{table:resultsMC} in the column $\kappa$. The solid line shows the fit and is given by \Eref{sigmaFRED}.}
\label{fig:sigma}
\end{center}
\end{figure}

A more careful analysis shows that the standard deviation for a FRED distribution does not behave like $\mathrm{const}/\sqrt{N}$. \Fref{fig:sigma} shows the standard deviation $\sigma_{\kappa}$ as a function of the photon number $N$ as given in \tref{table:resultsMC}. A fit to the data points between $N=500$ and $N=10000$ shows that the standard deviation of a FRED distribution is given by
\begin{equation}\label{sigmaFRED}
\sigma_{\kappa}=0.0182\cdot N^{-0.617},
\end{equation}
where the exponent is smaller than the usual $1/\sqrt{N}$ for a Gaussian. With this equation we are also able to assess the error for $\kappa$ when using data from GRBs measured by INTEGRAL.

\section{Results from GRBs detected by INTEGRAL}\label{sec:GRB}
\subsection{Determination of the parameter $\kappa$}\label{sec:kappa}

The data provided by INTEGRAL contains for each single registered photon four pieces of information: the arrival time, the energy, the dead time and the PIF value (see \sref{sec:INTEGRAL}). In our analysis we take only photons that have a PIF value larger than 0.9, i.e. we exclude pixels that are not completely open to the GRB flux. After correcting the arrival time by weighing it with $1/(1-\mathrm{dead}\;\mathrm{time})$, we determine from the light curve which time intervals have the shape of a FRED distribution. Recall from \Eref{sigmaFRED} that the more photons we take the more we are able to constrain $\kappa$.

In \cite{Bolmont:06} the average energy difference $\Delta\langle E\rangle=\Delta\langle E\rangle_3-\Delta\langle E\rangle_1$ was computed for each GRB using the energy bands of SWIFT, where $\Delta\langle E\rangle_3$ is the average energy of the photons with energies between 110 and 300 keV and $\Delta\langle E\rangle_1$ between 20 and 55 keV (see \tref{table:GRB}). In our variables the time difference would then be approximately given by $\Delta t=\kappa\cdot\Delta\langle E\rangle$. 

\begin{table}
    \caption{Results for the GRBs with known redshifts. Note that in spite of the fact that a couple of redshifts have error bars, we choose to take the mean value of the redshifts without errors. The reason is that then we don't have to introduce arbitrary error bars in order for exact redshifts not to be weighed infinitely strongly. $I(z)$ is given by \eref{Int}, $K(z)$ by \eref{Kofz}, $\kappa$ is the time lag per energy given by the maximization of \eref{maxlikelihood} with $\sigma_{\kappa}$ its error.}\label{table:GRB}\lineup
\begin{indented}
\item[]
\begin{tabular}{@{}llllll}
\br
GRB & $z$ & $I(z)$ & $K(z)$ & $\kappa$ [s/keV] & $\sigma_{\kappa}$\ [s/keV]   \\
\mr
030227 & 1.39 \cite{Watson:03}& 1.0 & 0.42 &  \m$7.8\cdot10^{-4}$ & $4.0\cdot10^{-5}$ \\
\hline
031203 & 0.11 \cite{Watson:04} &0.10&0.09&  $-1.7\cdot10^{-4}$ & $0.1\cdot10^{-4}$\\
\hline
&  & &  & \m$2.7\cdot10^{-4}$ & $3.7\cdot10^{-4}$ \\
\raisebox{1.5ex}[-1.5ex]{040106} &  \raisebox{1.5ex}[-1.5ex]{0.9 \cite{Moran:05}}& \raisebox{1.5ex}[-1.5ex]{0.73}& \raisebox{1.5ex}[-1.5ex]{0.38}&  $\m4.2\cdot10^{-4}$  & $1.5\cdot10^{-4}$   \\
\hline
040223 & 0.1 \cite{McGlynn:05} &0.1&0.09 & \m$2.5\cdot 10^{-3}$ & $0.4\cdot 10^{-3}$ \\
\hline
&  & & & $-1.4\cdot10^{-3}$ & $0.1\cdot10^{-3}$  \\
\raisebox{1.5ex}[-1.5ex]{040812} &  \raisebox{1.5ex}[-1.5ex]{0.5 \cite{D'Avanzo:06}}&  \raisebox{1.5ex}[-1.5ex]{0.45}&\raisebox{1.5ex}[-1.5ex]{0.3}&  \m$2.6\cdot10^{-4}$  & $0.8\cdot10^{-4}$   \\
\hline
040827 & 0.9 \cite{deLuca:05} &0.73 & 0.38 &  $-1.9\cdot10^{-4}$ & $5.9\cdot10^{-4}$  \\
\hline
 &  & & &$-1.2\cdot10^{-3}$ & $0.1\cdot10^{-3}$  \\
\raisebox{1.5ex}[-1.5ex]{041218} &  \raisebox{1.5ex}[-1.5ex]{0.8 \cite{Fatkhullin:05}}& \raisebox{1.5ex}[-1.5ex]{0.66} &  \raisebox{1.5ex}[-1.5ex]{0.37}&  \m$1.7\cdot10^{-3}$ & $0.2\cdot10^{-3}$  \\
\hline
 &  & & &$-2.2\cdot10^{-3}$ & $0.4\cdot10^{-3}$  \\
\raisebox{1.5ex}[-1.5ex]{050502} &  \raisebox{1.5ex}[-1.5ex]{3.8 \cite{Prochaska:05}}& \raisebox{1.5ex}[-1.5ex]{1.69} &  \raisebox{1.5ex}[-1.5ex]{0.35}&  \m$8.2\cdot10^{-4}$ & $1.9\cdot10^{-4}$  \\
\hline
050714 & 0.26 \cite{Prochaska:05:2} & 0.25 & 0.19  & $-8.9\cdot10^{-4}$ & $3.2\cdot10^{-4}$  \\
\hline
&  & & &$-1.4\cdot10^{-3}$ & $0.2\cdot10^{-3}$  \\
\raisebox{1.5ex}[-1.5ex]{050922} &  \raisebox{1.5ex}[-1.5ex]{2.17 \cite{Jakobsson:05}}& \raisebox{1.5ex}[-1.5ex]{1.3} &  \raisebox{1.5ex}[-1.5ex]{0.41}&\m $7.3\cdot10^{-4}$ & $0.2\cdot10^{-4}$ \\
\hline
& & & & \m$2.6\cdot10^{-4}$ & $6.3\cdot10^{-4}$ \\
\raisebox{1.5ex}[-1.5ex]{060204} & \raisebox{1.5ex}[-1.5ex]{3.1 \cite{Pelangeon:06}} & \raisebox{1.5ex}[-1.5ex]{1.55} &\raisebox{1.5ex}[-1.5ex]{0.38} &\m $2.3\cdot10^{-4}$ & $0.1\cdot10^{-4}$ \\
\br

\end{tabular}
\end{indented}
\end{table}

\begin{figure}[ht!]
\begin{center}
\includegraphics[width=6.4cm]{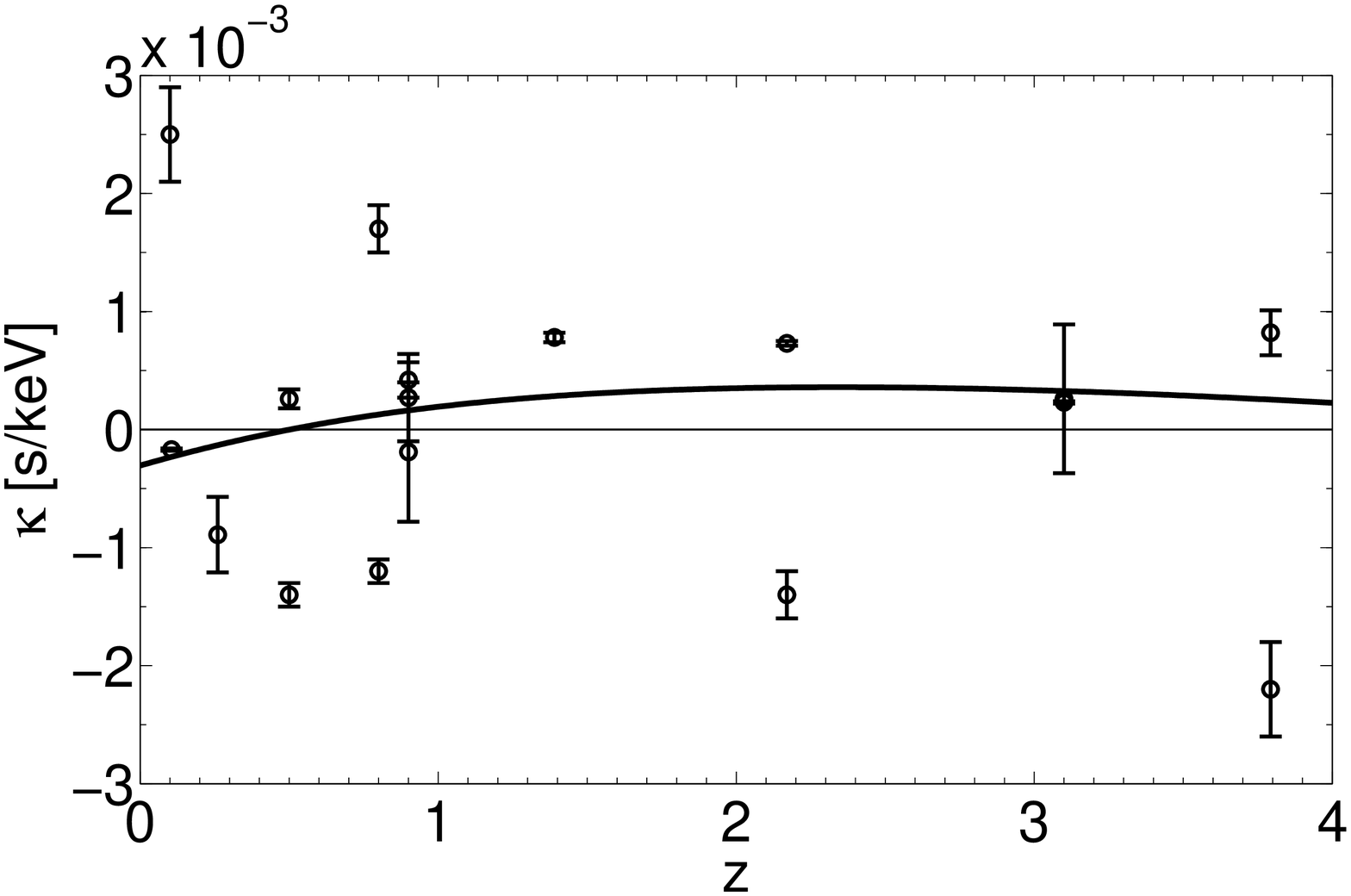}\includegraphics[width=6.2cm]{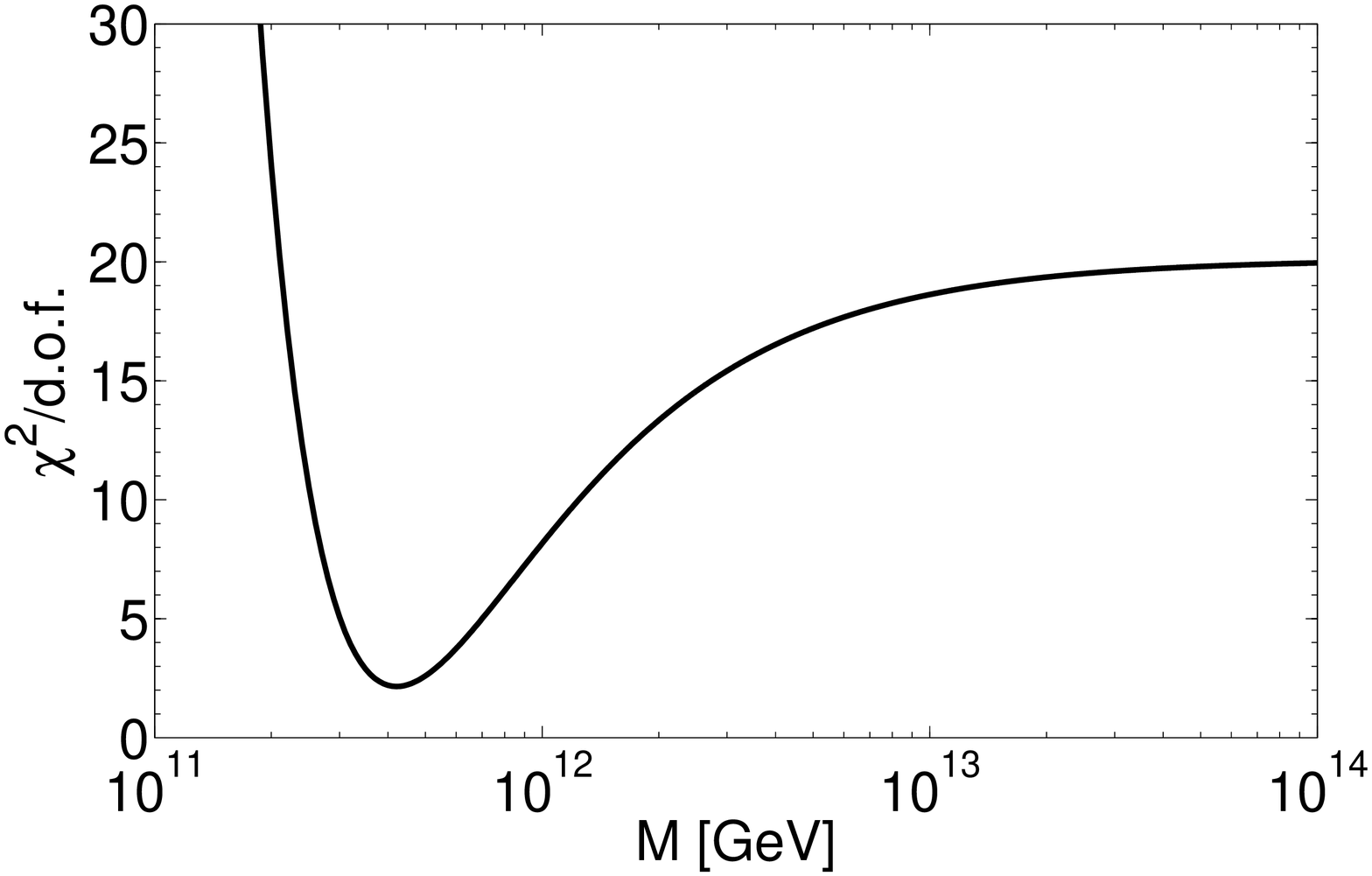}
\caption{Left panel: plot of $\kappa$ as a function of the redshift $z$ for several GRBs detected by INTEGRAL. The nonlinear fit is given by \eref{fiteq}, the maximum lies at $z_{\mathrm{max}}\simeq2.34$ with value $\kappa_{\mathrm{max}}\simeq3.6\cdot10^{-4}$ s/keV. Right panel: evolution of the $\chi^2$ function as a function of $M$. Note that $\chi^2$ has a strong minimum around $4\cdot10^{11}$ GeV.}
\label{fig:plotresult}
\end{center}
\end{figure}

However, from our analysis we obtain the parameter $\kappa$ directly so we do not average over energies in order to get a time difference. Considering only a linear approximation to quantum gravitational effects as proposed by Ellis \etal \cite{Ellis:03,Ellis:06}, we have the relation
\begin{equation}\label{Klinear}
	\kappa=aI(z)+b(1+z),
\end{equation}
where $a$ and $b$ are coefficients to be fitted. The constant $b$ parameterizes time lags in the rest frame of the source caused by unknown internal processes of the GRBs. Comparing \eref{Klinear} with \eref{deltatth} we find that $I(z)$ is given by
\begin{equation}\label{Int}
	I(z)=\int_0^z\frac{dz}{\sqrt{\Omega_{\Lambda}+\Omega_m(1+z)^3}}
\end{equation}
and the QG parameter $a$ by
\begin{equation}
	a=\pm\frac{H_0^{-1}}{Mc^2}.
\end{equation}
Fitting all GRBs with known redshift detected by INTEGRAL, we find (units s/keV are used)
\begin{equation}\label{fiteq}
	\kappa=(9.5\pm 3.0)\cdot10^{-4}\cdot I(z) -(2.8\pm 1.1)\cdot10^{-4}\cdot(1+z)
\end{equation}
as shown in the left panel of \fref{fig:plotresult}. Because redshifts are measured without using a specific cosmological model, this fit was obtained using data that are model-independent. Moreover, a rather questionable energy binning as explained above is not needed due to the fact that our analysis method yields directly values for $\kappa$. 

As can be seen from \fref{fig:plotresult} (left panel), a single GRB with two bursts can lead to very different time lags. For example, GRB040812 with average redshift $z=0.5$ has two peaks that even differ in the sign: the first one has a negative value $\kappa=-1.4\cdot10^{-3}$ and the second one a positive value $\kappa=2.6\cdot10^{-4}$. This could be explained by the fact that different internal processes are at the origin of the two bursts, which implies that it may not be sufficient to describe internal time lags with a constant $b$ as in \Eref{Klinear}. However, the physics involved in GRB is still not well understood, thus limiting the possibility to model intrinsic effects in other ways than through \Eref{Klinear}.

In \cite{Ellis:06} and \cite{Bolmont:06} a linear fit was obtained by using not $z$ as the independent variable but instead a function $K(z)$. Dividing \eref{Klinear} by $(1+z)$, $K(z)$ is given by the non-linear function
\begin{equation}\label{Kofz}
K(z)\equiv\frac{1}{1+z}I(z).
\end{equation}

However, we think that considering $\kappa/(1+z)$ as a linear function of $K$ is delusive because the new function $K(z)$ is not injective. This function maps certain different redshifts $z$ to the same value and has a maximum of $K_{\mathrm{max}}\simeq0.42$ at $z=1.64$. For example, a redshift of $z=4$ has the same value $K$ as a redshift of $z=0.7$. Thus the two points for GRB050502 at $z=3.793$ are mapped to $K=0.353$, which is between GRB040812 and GRB040106. Our opinion is that this method is misleading and does not give reliable results and should therefore not be used.

\subsection{Likelihood test}\label{sec:likelihood}

Following \cite{Ellis:06}, we introduce a likelihood function
\begin{equation}\label{L}
	L_{\mathrm{LH}}(M)=\mathcal{N}\exp \left(-\frac{\chi^2(M)}{2}\right),
\end{equation}
where $M$ is the mass scale, $\mathcal{N}$ the normalization and $\chi^2(M)$  is given by
\begin{equation}\label{chisquare}
	\chi^2(M)=\sum_{\mathrm{all}\,\mathrm{GRBs}}\frac{\left(\kappa_i-b(1+z_i)-a(M)I_{i}\right)^2}{(\sigma_i)^2+\sigma_b^2}.
\end{equation}
The parameter $b$ reflects the instrinsic time lags and $a$ quantum gravitational effects. Thus, $b$ was removed from the linear fit, as can be seen from \eref{chisquare}. Note that we used the raw model that doesn't need an energy binning.

The value at the minimum of $\chi^2/\mathrm{d.o.f.}$ is 303/15, which is well above unity. In such a case, we may expect a high degree of uncertainty for any fitted parameters. If the error bars are underestimated it will lead to underestimated statistical errors for the fitted parameters. In such cases, the Particle Data Group \cite{pdg:06} suggests to rescale the error bars so that $\chi^2\approx\mathrm{d.o.f.}$ by a factor $S=[\chi^2/\mathrm{d.o.f.}]^{1/2}$. Such a rescaling has also been proposed in \cite{Ellis:03,Ellis:06,Bolmont:06}

\Fref{fig:plotresult} (right panel) presents the dependence of the rescaled $\chi^2/\mathrm{d.o.f}$ as a function of $M$ . The minimum of this function is found at $M\simeq3.8\cdot10^{11}$ GeV. This value also minimizes the likelihood function given by \Eref{maxlikelihood}.

Following Ellis \etal \cite{Ellis:03} we establish a 95 \% confidence-level lower limit on the scale $M$ of quantum gravity by solving the equation
\begin{equation}
\frac{\int_M^{\mathcal{M}} L_{\mathrm{LH}}(\xi)d\xi}{\int_0^{\mathcal{M}} L_{\mathrm{LH}}(\xi)d\xi}=0.95,
\end{equation}
where the Planck mass $\mathcal{M}=10^{19}$ GeV is the reference point fixing the normalization. The function $L_{\mathrm{LH}}$ is given by \Eref{L}.
Solving this equation for $M$ gives the lower limit of quantum gravity at a 95 \% level of confidence at
\begin{equation}\label{M}
M\geqslant 3.2\cdot10^{11}\,\mathrm{GeV}.
\end{equation}

\section{\label{sec:conclusion}Conclusions}
In this work, we first described a method that is able to analyze unbinned data of GRBs detected by INTEGRAL. We introduced a maximum likelihood function following a Fast Raise and Exponential Decay behavior with a parameter describing time lags of photons for different energies. In order to know which minimum time lags are measurable with INTEGRAL, we performed Monte Carlo simulations and varied the total photon number. 

We had 11 GRBs with known redshift at our disposal and were able to get 17 measurements of time lags. We used these measurements to fit a nonlinear relation depending on the redshift. This relation has a term that describes possible quantum gravitational effects and one that accounts for intrinsic time lags of the GRB. By using a likelihood function we made a $\chi^2$ analysis of the data and showed that there is a strong minimum of $\chi^2$ around $4\cdot10^{11}$ GeV, which apparently would disfavor a quantum gravitational scale around the Planck mass. However, as shown by our Monte Carlo simulations in \sref{sec:MC} it is obvious that it is impossible to obtain the required sensitivity with the presently available statistics of GRB data, especially when only 11 GRBs are at disposal. Correcting for intrisinc time lags \cite{Fenimore:95,Norris:96,Norris:00} dramatically increases this lower bound to $1.5\cdot10^{14}$ GeV, but this method stands on shaky ground.

A better precision in time could be achieved by constructing satellites with a much larger collecting surface. However, as shown in \sref{sec:MC} even with the unrealistic photon number of $3\cdot10^5$ for a single burst the time resolution is still two orders of magnitude too low. The other solution is to increase the photon energy, as can be seen from \Eref{deltatth}. A more complete strategy how to reach the strongest possible bounds was discussed in \cite{Rodriguez:06:01,Piran:04}. GLAST will be able to increase the time resolution by several orders of magnitude as it will be able to detect photons up to energies of 20 MeV. As the time lags are linearly dependent on the energy difference GLAST should be able to improve the time difference by a couple of orders of magnitude, which may be even larger than the expected time difference caused by QG effects at the Planck scale.

\ack
In this work we used data from INTEGRAL publicly available. We thank S. Hossenfelder, T. Kahniashvili and T. Piran for drawing our attention to some important references.

\end{document}